\documentclass[RNAAS]{aastex62}

\graphicspath{{./}{figures/}}

\begin{document}

\title{Spectropolarimetry of the WR + O Binary WR42}

\correspondingauthor{Andrew G. Fullard}
\email{andrew.fullard@du.edu}

\author{Andrew G. Fullard}
\affiliation{University of Denver}

\author{Jennifer L. Hoffman}
\affiliation{University of Denver}

\author{Sophia DeKlotz}
\affiliation{University of Denver}

\author{Daniel Azancot Luchtan}
\affiliation{University of Denver}

\author{Kevin Cooper}
\affiliation{University of Denver}

\author{Kenneth H. Nordsieck}
\affiliation{University of Wisconsin, Madison}



\section{Introduction} 
Wolf-Rayet (WR) stars are massive stars shrouded in optically thick winds, characterized by strong emission lines \citep{crowther_physical_2007}. When a WR star has an O-type binary companion, the winds of both stars interact, leading to mass transfer between the stars. This can ``spin up" the O star and potentially create the progenitor of a long-duration gamma-ray burst (LGRB). Spectropolarimetry illuminates the mass transfer in a binary system by constraining the location of the emitting and scattering regions. 

WR 42 is a WC7 star in a 7.8823-day orbit with an O7V companion \citep{davies-orbitwr42_1981}. It was first studied polarimetrically by \citet{st.-louis_polarization_1987}, who investigated the wind properties assuming symmetry consistent with models by \citet{brown_polarisation_1977}. Light curves from \citet{lamontagne_photometric_1996} show a minor primary eclipse but no significant secondary eclipse; these authors established a system inclination angle of 38--44$\degr$, so only the atmosphere of the WR star likely occults the O star around phase 0.5.  More recently, \citet{hill_modelling_2000} 
proposed a cone-shaped wind interaction region, while \citet{heuvel_forming_2017} predicted that the system could evolve to a double black hole. \citet{shara_spin_2017} found a high rotational velocity of 
$452-511$ km s$^{-1}$ for the O star in the system. Thus WR 42 is an excellent candidate for the investigation of LRGB progenitors.


\section{Observations}
We present new spectropolarimetric data for WR 42 collected over 6 months at the 11-m Southern African Large Telescope (SALT) using the Robert Stobie Spectrograph (RSS) \citep{kobulnicky_prime_2003,buckley_completion_2006}. Observations were conducted in the long-slit spectropolarimetry mode with 80 s exposures and a spectral resolution of 1 \AA~over the range 4200--7260 \AA. We obtained 10 observations between 11 Dec 2017 and 12 May 2018, and discarded one of these due to a 
software crash in a mirror segment controller. One observation, at phase 0.272, saturated in the \ion{He}{2} $\lambda4686$ line.

\section{Discussion}

\begin{figure}[ht!]
\begin{center}
\includegraphics[scale=0.5,angle=0]{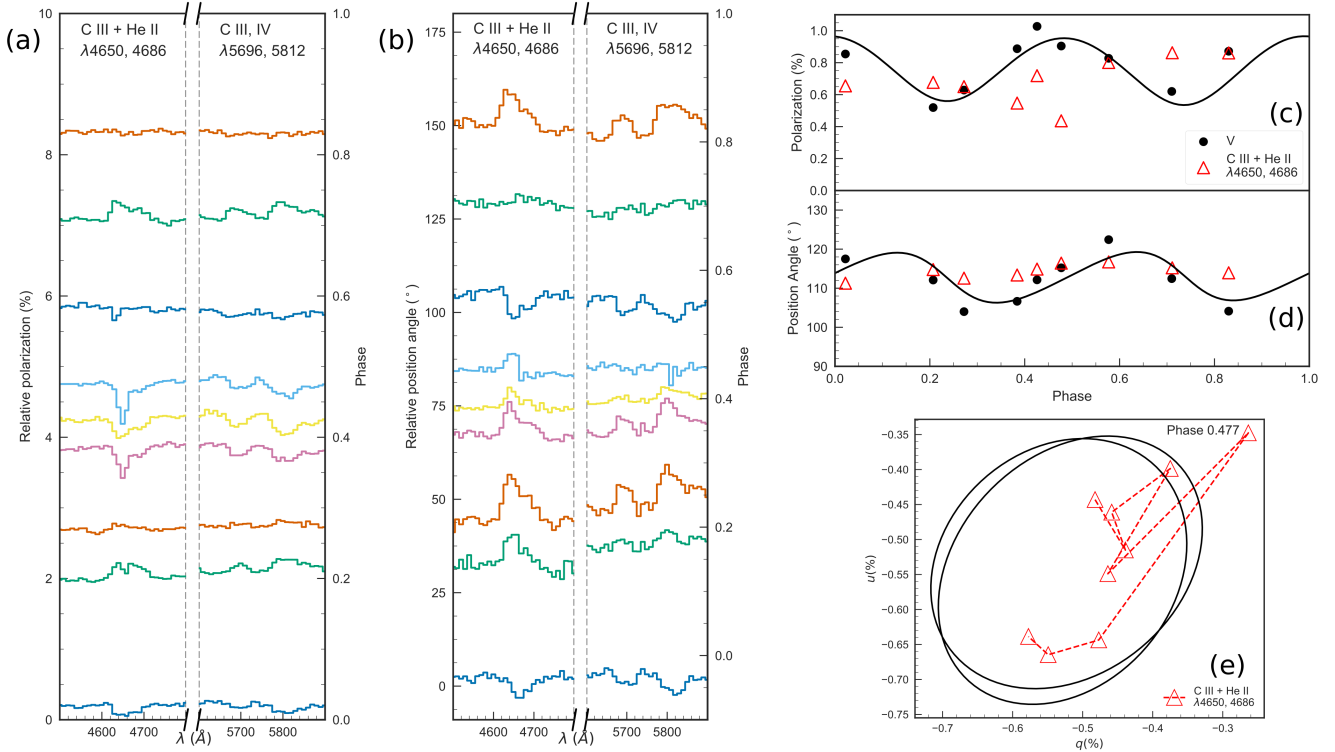}
\caption{(\textit{a and b}) Line profiles from RSS spectropolarimetry of WR 42, binned to 10 \AA. Spectra have been vertically separated by phase; the phase axis aligns with the average polarization or position angle of each observation. Error bars are approximately the thickness of the lines.  (\textit{c and d}) Polarization and position angle versus phase, respectively, of our data compared with the broadband Fourier fit of \citet[\textit{solid line}]{st.-louis_polarization_1987}. 
Error bars are smaller than the symbols. ($e$) \ion{C}{3} + \ion{He}{2} line polarization in $q$--$u$ space, compared with the Fourier fit. Error bars are smaller than the symbols.
\label{fig:1}}
\end{center}
\end{figure}

We calculated orbital phases using the ephemeris from \citet{davies-orbitwr42_1981}. The \ion{C}{3}+\ion{He}{2} $\lambda$4650/4686 and \ion{C}{3}/\ion{C}{4} $\lambda$5696/5812 lines show complex phase-dependent and velocity-dependent polarization and position angle behavior, particularly across the atmospheric eclipse at phase 0.5 (Fig. \ref{fig:1} $a$, $b$).

We extracted broadband polarimetry by convolving a synthetic Johnson $V$ filter with the data, excluding the \ion{C}{3}/\ion{C}{4} lines by zeroing the filter response between $5550-5900$ \AA. Our $V$-band polarimetry displays sinusoidal behavior similar to that reported by \citet{st.-louis_polarization_1987}, though shifted by $-0.2$ in phase (consistent with their reported uncertainties; Fig. \ref{fig:1} $c$, $d$).

We also calculated intrinsic polarization in the compound \ion{C}{3} + \ion{He}{2} $\lambda$4650/4686 line via the \textit{pfew} method \citep{Lomax_2015}. This line maintains a near-constant position angle with phase, but its polarization decreases significantly around phase 0.5 (Fig. \ref{fig:1} $c$, $d$).

We present the $V$-band and \ion{C}{3}/\ion{C}{4} $\lambda$5696/5812 results in $q$--$u$ space in Fig. \ref{fig:1} $e$ along with the \citet{st.-louis_polarization_1987} solution. The top-right point is at phase 0.477, showing the drop in line polarization during atmospheric eclipse.

The sinusoidal continuum polarization variation is likely caused by the aspherical WR wind  occulting the O star. The line polarization does not follow the sinusoidal phase behavior, showing that the line forms and scatters differently than the continuum. 

These data are part of an ongoing project to study the colliding winds of WR + O binaries with spectropolarimetry. We anticipate obtaining more observations of WR 42 with RSS in the coming months.

\acknowledgments
We thank M. Shrestha and the SALT observing team for their assistance.


\bibliography{WR42_Research_Note.bib}
\bibliographystyle{aasjournal.bst}

\end{document}